\documentclass[structabstract]{aa}
\pdfoutput=1
\usepackage[pdftex]{graphicx}
\usepackage{natbib,amssymb,url}
\usepackage{hyperref}
\usepackage{color}

\newcommand{\plotwd}{9.cm}
\newcommand{\plotwdtwo}{18.cm}

\begin{document}
\title{Angular fluctuations in the CXB: Is Fe 6.4 keV line tomography of the large-scale structure feasible?}
\titlerunning{Is Fe 6.4 keV line tomography of LSS feasible?}
\author{Gert H\"utsi \inst{1,2}, Marat Gilfanov \inst{1,3}, Rashid Sunyaev \inst{1,3}}
\institute{Max-Planck-Institut f\"ur Astrophysik, Karl-Schwarzschild-Str. 1, 85741 Garching, Germany \\ \email{gert@mpa-garching.mpg.de} \and Tartu Observatory, T\~oravere 61602, Estonia \and Space Research Institute of Russian Academy of Sciences, Profsoyuznaya 84/32, 117997 Moscow, Russia}
\date{Received / Accepted}

\abstract
{AGN are known to account for a major portion, if not all, of the cosmic X-ray background radiation.  The dominant sharp spectral feature in their spectra is the 6.4 keV fluorescent line of iron, which may contribute to as much as $\sim 5-10\%$ of the CXB spectral intensity at $\sim 2-6$ keV. Owing to cosmological redshift, the line photons detected at the energy $E$ carry information about objects located at the redshift $z=6.4/E-1$. In particular, imprinted in their angular fluctuations is the information about the large-scale structure at redshift $z$. This opens the possibility of performing the Fe K$_\alpha$ line tomography of the cosmic large-scale structure.
}
{The goal of this paper is to  investigate the feasibility of the Fe K$\alpha$ line tomography of the large-scale structure.}
{At any observed energy $E$, the 6.4 keV line photons are blended with continuum emission, which originates in objects located at many different redshifts and therefore contaminates and dilutes the tomographic signal. However, its contribution can be removed by doing observations at two nearby energy intervals and by calculating the power spectrum of the corresponding differential signal map.
}
{We show that detection of the tomographic signal at $\ga 100\sigma$ confidence  requires an  all-sky  survey by an instrument with an effective area of $\sim 10$ m$^2$ and field of view  of $\sim 1\deg^2$.
The signal is strongest for objects located at the redshift $z\sim 1$ and at the angular scales corresponding to $\ell \sim 100-300$, therefore an optimal detection can be achieved with an instrument having a rather modest angular resolution of $\sim 0.1-0.5\deg$. For such an instrument, the CCD-type energy resolution of $\sim 100-200$ eV FWHM is entirely sufficient for the optimal separation of the signals coming from different redshifts.  The gain in the signal strength that could  potentially be  achieved with energy resolution comparable to the line width is nullified by the photon counting and AGN discreteness noise. Among the currently planned and proposed missions, these requirements are best satisfied by LOFT, even though that it was proposed for an entirely different purpose. Among others, clear detection should be achieved by WFXT ($\sim 25-40\sigma$) and ATHENA ($\sim 20-30\sigma$).}
{}
\maketitle

\section{Introduction}\label{sec1}
Since the discovery of the cosmic X-ray background (CXB) back in 1962 \citep{1962PhRvL...9..439G}, the performance of X-ray instrumentation has witnessed a very rapid evolution where sensitivities and angular resolutions of modern X-ray observatories are higher by more than nine and around five orders of magnitude, respectively \citep{2003RvMP...75..995G}. With the aid of XMM-Newton~\footnote{\url{http://xmm.esac.esa.int}} and Chandra~\footnote{\url{http://chandra.harvard.edu}} X-ray observatories, two of the most advanced X-ray instruments in existence, more than 20 deep extragalactic X-ray surveys over varying sky areas and effective flux limits have been performed \citep{2005ARA&A..43..827B}. The most noticeable amongst these are $\sim 2$ Ms Chandra Deep Field North (CDF-N) \citep{2003AJ....126..539A} over $\simeq 448$ arcmin$^2$ sky area, $\sim 4$ Ms Chandra Deep Field South (CDF-S) \citep{2011ApJS..195...10X} over $\simeq 465$ arcmin$^2$, and $\sim 3$ Ms XMM-Newton deep survey in the CDF-S \citep{2011A&A...526L...9C}.  In the energy ranges $0.5-2$ keV and $2-8$ keV, up to $76\%$ and $82\%$ of CXB has been currently resolved \citep{2012ApJ...752...46L}. 
These observations have demonstrated that, although contribution by faint star-forming galaxies to the source counts becomes important at the flux limit of the deepest Chandra surveys, active galactic nuclei (AGN) dominate source counts at brighter fluxes and  produce the dominant fraction of  the CXB intensity \citetext{e.g., \citealp{2007A&A...463...79G}}, with only relatively minor contributions from galaxy clusters and from X-ray binaries in star-forming galaxies \citep{2012MNRAS.421..213D}. Along with earlier studies, this has lead to the general belief that CXB is fully accounted for by spatially sparse X-ray emitting sources and thus placing stringent constraints on the possible existence of a genuinely diffuse component.

Despite the small areas of the above deep surveys, the large number of detected AGN~\footnote{In fact, X-ray selection is currently the most effective way of selecting AGN. The deepest optical spectroscopic surveys typically give a factor of $\sim 10$ times less AGN per deg$^{-2}$, and only utradeep optical variability studies are able to generate comparable AGN sky densities \citetext{e.g., \citealp{2005ARA&A..43..827B}}.} has allowed relatively good determination of the AGN luminosity function (LF) and its evolution over cosmic time. While the very narrow and deep surveys are not very suitable for measuring the spatial clustering properties of AGN, somewhat wider and shallower surveys like X-Bo\"otes, XMM-LSS, AEGIS, and XMM-COSMOS have allowed measurement of the two-point clustering statistics \citep{2005ApJS..161....1M,2006A&A...457..393G,2009ApJ...701.1484C,2009A&A...494...33G}, albeit with relatively large uncertainty on large scales. The best large-scale X-ray AGN clustering measurements to date are obtainable with ROSAT-based surveys, e.g., \citet{2010ApJ...713..558K,2011ApJ...726...83M}. This situation is expected to improve significantly with the upcoming Spectrum-X-Gamma/eROSITA~\footnote{\url{http://www.mpe.mpg.de/erosita/}}~\footnote{\url{http://hea.iki.rssi.ru/SRG/}} space mission, which is planned to cover entire sky to the limiting sensitivity of $\sim 10^{-14}$ erg/s/cm$^2$ \citetext{e.g., \citealp{2010SPIE.7732E..23P,kolodzig}}. Also, the spectacular capabilities of the proposed future X-ray observatories, e.g. ATHENA~\footnote{\url{http://www.mpe.mpg.de/athena/}} and WFXT~\footnote{\url{http://wfxt.pha.jhu.edu/}} proposals, are expected to provide a significant boost in our ability to observe AGN clustering with high accuracy.

In this paper we investigate fluctuations in the CXB intensity field, i.e., we do not look at AGN clustering in the ``usual sense'', but instead include all the photons constituting the CXB. On small scales (relevant to the above-mentioned deep X-ray surveys) intensity fluctuations, which are dominated by an uncorrelated Poisson process sourced by the discrete and sparse spatial distribution of AGN, have been successfully used to extrapolate the observed number count and limiting flux relation below the survey's point source detection limit \citep{2002ApJ...564L...5M}. Although this small-scale fluctuation term is included in our study, the focus of the current paper is on CXB fluctuations on larger spatial scales, where the correlated nature of the fluctuations becomes noticeable. In particular, as the main topic of this study we investigate the possibility of doing iron $6.4$ keV line~\footnote{Fe $6.4$ keV line is the prominent fluorescence line in the AGN spectra thought to arise from the reflection off of the cold accretion disk and molecular torus component.} tomography in a way similar to the well-known neutral hydrogen 21cm tomography in the radio band (see \citet{2006PhR...433..181F,2011arXiv1109.6012P} for extensive reviews). Even though in comparison to the radio band, we are typically quite severely limited by the poor photon statistics at X-ray frequencies and by the spatial sparseness of the X-ray emitting sources, these difficulties are somewhat compensated by significantly lower level of possible contaminants in the form of various back- or foregrounds. The CXB at energies $2-10$ keV is basically only due to AGN \citep{2007A&A...463...79G} for which the  Fe K$\alpha$ fluorescent emission line at $6.4$ keV is the dominant sharp spectral feature in this energy range. Although in  unobscured AGN, its strongest component is relativistically broadened \citep{2000PASP..112.1145F} and often redshifted,  the narrow Fe K$\alpha$ line at $6.4$ is almost always  present: it is known to be a common feature in the X-ray spectra of local \citep{2010ApJS..187..581S,2011ApJ...738..147S} and distant \citetext{e.g., \citealp{2012A&A...537A...6C}} AGN. Its equivalent width depends on the amount of the intrinsic absorption in the AGN spectrum, increasing from $\sim 50-100$ eV in unabsorbed \citep{2010ApJS..187..581S}, type I AGN to $\sim 1-2$ keV in Compton thick objects \citep{2011ApJ...738..147S}. The intrinsic width of the narrow  iron line was reported to be the same in unabsorbed, $2200 \pm 220$ km/s FWHM, and absorbed, $2000 \pm 160$ km/s FWHM \citep{2010ApJS..187..581S,2011ApJ...738..147S} objects, corresponding to $\sim 40-45$ eV FWHM. Apart from the X-ray Baldwin effect of a moderate amplitude \citetext{e.g., \citealp{1993ApJ...413L..15I,2012A&A...537A...6C}}, no other obvious trends have been reported in the line parameters with the redshift and/or luminosity.

With deep Chandra and XMM-Newton surveys, it has been established that the fraction of moderately obscured, Compton-thin AGN is on average $3/4$ of all AGN; it is higher at lower luminosities \citetext{e.g., \citealp{2003ApJ...598..886U,2005ApJ...630..115T}} and higher redshifts \citetext{e.g., \citealp{2005ApJ...635..864L}}. Although the fraction of Compton-thick objects is not so well constrained by observations, it is believed to be comparable to that of obscured Compton-thin objects (see, e.g., \citet{2007A&A...463...79G} and references therein). In agreement with these figures, CXB spectral synthesis calculations have shown that its major, $\sim 2/3$,  fraction is composed of emission of obscured objects. In these objects, a notable fraction, up to $\sim 2-20\%$, of $2-10$ keV luminosity is carried away by narrow line photons, which make a correspondingly sizable contribution to the cosmic X-ray background. \citet{1999NewA....4...45G} were the first to calculate this contribution and obtained $\sim 7\%$ at the energy of a few keV. This is a strong signal when compared to the typical 21cm signal in relation to the Galactic ($\sim 10^{-5}-10^{-4}$) or extragalactic foregrounds ($\sim 10^{-2}-10^{-1}$) \citep{2006PhR...433..181F}.

We also mention that several other spectral lines have been investigated as potential tools for performing tomographic measurements of the large-scale structure (LSS) via intensity mapping: e.g., rotational transitions of the CO molecule \citep{2008A&A...489..489R,2011ApJ...730L..30C,2011ApJ...741...70L}, $^3$He\hspace{1pt}II hyperfine transition \citep{2009PhRvD..80f3010M}, C\hspace{1pt}II fine structure line \citep{2012ApJ...745...49G}, hydrogen Lyman-$\alpha$ line \citep{2012arXiv1205.1493S}.  Also, one has to point out that Fe $6.4$ keV line has already proven to be a powerful tomographic probe of the accretion disk and central supermassive black hole (see \citet{2000PASP..112.1145F} for a review). In this paper we investigate its potential as a tomographic probe of the LSS.  

In the energy range relevant to this study we approximate the typical AGN spectrum with a power-law continuum plus a single Gaussian line at $6.4$ keV. The line photons, emitted by AGN located at the redshift $z$, will be observed at the energy $E=\frac{6.4}{1+z}$ keV. As AGN from a broad range of redshifts contribute to the X-ray background, the combined iron line emission from many objects will produce a broad hump on the CXB spectrum, without any sharp features\citep{1999NewA....4...45G}. However, the line photons observed in a narrow energy interval $\Delta E$ centered at energy $E$ will carry information about correlation properties of AGN in the redshift shell $z\simeq \frac{6.4}{E+\Delta E}-1\ldots\frac{6.4}{E}-1$. This information is diluted by the continuum photons, which at any given energy are produced by objects located at many different redshifts. Since the line photons cannot be directly separated from continuum photons, statistical methods need to be employed in order to subtract contribution of the continuum and to extract the information about correlation properties of the AGN at a given redshift. Here one can take advantage of the different behavior as a function of energy of those two components: continuum is changing more slowly than the line contribution.

Our paper is organized as follows. In Section~\ref{sec2} we present a simple model for the CXB fluctuations including $6.4$ keV Fe line. There we model auto and cross power spectra as observed in narrow energy ranges and combine these in a way to enhance line (and suppress continuum) contribution. In Section~\ref{sec3} we calculate the expected tomographic signal strength as a function of position and width of the observational energy bands along with its dependence on the limiting flux above which resolved sources are removed. We also discuss the prospects for the current and future X-ray instruments to measure the signal. In Section~\ref{sec4}, in addition to bringing our conclusions, we discuss several issues that were neglected in the main analysis and point toward the ways of extending the above work. 

Throughout this paper we assume a flat $\Lambda$CDM cosmology with $\Omega_m=0.27$, $\Omega_b=0.045$, $h=0.70$ and $\sigma_8=0.8$.

\section{A simple model for the CXB fluctuations including Fe 6.4 keV line}\label{sec2}
In this section we introduce models for the auto and cross power spectra. We present a scheme for separating the line signal from the dominant continuum contribution and discuss the error estimates for the two-point functions.

In calculating the power spectra throughout this section, we assumed the limiting flux $F_{\rm lim}=10^{-13}$ erg/s/cm$^2$; i.e., all the sources brighter than this flux were removed from the analysis. We also assumed that the instrumental sensitivity and exposure time of the survey are such that the flux $10^{-13}$ erg/s/cm$^2$ corresponds to $500$ counts in $2-10$ keV band. This sensitivity is, for example, achieved in a $\sim 35$ ($\sim 25$) ksec XMM-Newton observation with PN (PN+2MOS) detector, assuming photon index $\Gamma=2$.

\subsection{Auto and cross power spectra}
These Fourier-space two-point functions carry the full statistical information in case the underlying random fields obey Gaussian statistics. In our case, this assumption turns out to be quite good, since e.g., the accuracy for determining the global amplitude of the line signal is dominated by intermediate scales, where the approximation of Gaussianity is very justified.

In order to calculate CXB two-point functions we need to know
\begin{enumerate}
\item the number density of AGN as a function of luminosity and redshift, i.e., AGN LF;
\item the AGN clustering bias;
\item the spectral shape of the typical AGN along with its $6.4$ keV iron line width and strength.
\end{enumerate}
As mentioned in the Introduction the $6.4$ keV line is the dominant line in the spectra of objects contributing to the CXB as long as one looks at energies $E\gtrsim 2-3$ keV, while lines from hot ISM and IGM in galaxies and clusters of galaxies start to complicate this simple picture at lower energies. In this paper we have chosen to constrain the observational energies above $2$ kev and consider only the narrow $6.4$ keV line from AGN. For the AGN LF we adopt $2-10$ keV band LF as determined by \citet{2010MNRAS.401.2531A}. 

We use two approaches to compute angular power spectra. In the first approach  we use the analytic luminosity-dependent density evolution (LDDE) fit to the observationally determined LF from \citet{2010MNRAS.401.2531A} and assume that all the AGN populate DM halos with the effective mass of $M_{\rm eff}=10^{13}$ h$^{-1}M_{\odot}$, which is compatible with the results of \citet{2011ApJ...736...99A}, and the corresponding bias is taken from the analytic model of \citet{2001MNRAS.323....1S}.

\begin{figure}
\centering
\includegraphics[width=\plotwd]{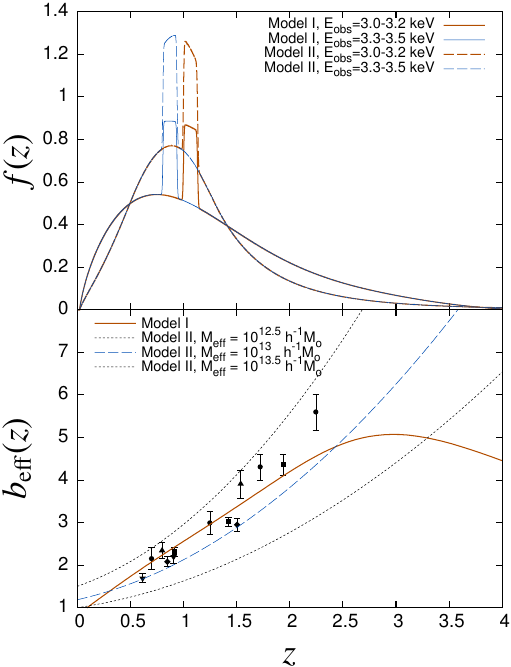}
\caption{{\bf Upper panel:} Radial selection functions, i.e., probability distribution functions for emission redshifts of photons, for Models I and II (see Table~\ref{tab1}), assuming observational energy bands $3.0-3.2$ keV and $3.3-3.5$ keV with $F_{\rm lim}$ fixed to $10^{-13}$ erg/s/cm$^2$. {\bf Lower panel:} Effective clustering bias as a function of redshift for Models I and II. For Model II, along with our default value $M_{\rm eff}=10^{13}$ h$^{-1}M_{\odot}$, we also show cases with $M_{\rm eff}=10^{12.5}$ and $10^{13.5}$, which cover the typical range of biasing values obtained in the literature. The points with error bars show bias measurements for X-ray selected AGN from the COSMOS field as determined by \citet{2011ApJ...736...99A} (see their Fig. 9 for further details).}
\label{fig1}
\end{figure}

In the second approach, we have fitted LF data from \citet{2010MNRAS.401.2531A} with a simple model where the concordance $\Lambda$CDM model halo mass function (MF) given in the analytical form of \citet{1999MNRAS.308..119S} is mapped to the LF. Here the mapping between the horizontal luminosity and mass axes is obtained by assuming
\begin{equation}
M(L,z)=M_{\min}\left(\frac{L}{L_{\min}}\right)^{c_1z+c_2}\,,
\end{equation}
where $L_{\min}=10^{41}$ erg/s and $M_{\min}$, $c_1$, $c_2$ are free parameters. The acceptable mapping between vertical axes is obtained if one assumes the ``duty cycle'' in the form
\begin{equation}
f_{\rm duty}(z)=\exp\left(c_3z^2+c_4z+c_5\right)\,.
\end{equation}
This MF to LF mapping implicitly assumes that there is one supermassive black hole (SMBH) for each dark matter (DM) halo, which is turned on (i.e., shines as an AGN) and off, as determined by the above duty cycle, and its luminosity during the `on' state is determined by the DM halo mass. Thus, in this case the LF is fitted by the following form
\begin{equation}\label{eq3}
\frac{{\rm d}n}{{\rm d}L}(L,z)=f_{\rm duty}(z)\frac{{\rm d}M}{{\rm d}L}(L,z)\frac{{\rm d}n}{{\rm d}M}\left[M(L,z),z\right]\,.
\end{equation}
It turns out that this simple six-parameter ($M_{\min}$, $c_1-c_5$) model gives completely satisfactory fit to the \citet{2010MNRAS.401.2531A} LF data. The advantage of this model is that once the MF to LF mapping is done, we have an automatic prediction for the bias parameters needed in clustering calculations. As it turns out, the bias values as a function of redshift, which are shown in the lower panel of Fig.~\ref{fig1}, are in reasonable agreement with available AGN clustering measurements (see, e.g., \citet{2011ApJ...736...99A,2012AdAst2012E..25C}). As an example, in Fig.~\ref{fig1} the points with error bars show bias measurements for X-ray-selected AGN from the COSMOS field as determined by \citet{2011ApJ...736...99A}. One can see that both models are indeed in reasonable agreement with observational measurements and that, in the redshift range $z\sim 0-2$ most important for this study, Model I performs somewhat better. The details, along with several other consequences of Model I, will be presented in a separate paper \citep{articleinprep}. The properties of these two models are briefly  summarized  in Table~\ref{tab1}.

\begin{table}
\centering
\caption{Main characteristics of the two models used in this paper.}
\label{tab1}
\begin{tabular}{l|l|l|}
Model & luminosity function & clustering bias \\
\hline
Model I &  given by MF to LF & follows automatically \\
 & mapping (Eq.~(\ref{eq3})) & \\
Model II &  \citet{2010MNRAS.401.2531A} & bias corresponding to \\
 & LDDE model & $M_{\rm eff}=10^{13}$ h$^{-1}M_{\odot}$
\end{tabular}
\end{table}

As a final ingredient, we need some model for the typical AGN spectrum with its Fe $6.4$ keV line. For simplicity, we assume the following spectral template~\footnote{This is clearly quite a substantial approximation. The possible consequences of the oversimplification is discussed further in Section~\ref{sec4}.}
\begin{itemize}
\item power-law continuum with spectral index $\Gamma=2$, i.e., $\widetilde{L}_{E,\,\rm cont}\propto E^{-\Gamma}$ ;
\item Gaussian Fe $6.4$ keV line with width $\sigma=45$ eV FWHM and equivalent width $EW=300$ eV.
\end{itemize}

As detailed spectral synthesis of the cosmic X-ray background is beyond the scope of this paper, we simplify our calculations and assume the above spectrum for all objects. Although with this assumption, our model will not reproduce exact shape of the CXB spectrum, it is sufficient for our goal -- to compute angular correlations in the CXB brightness distribution, because no noticeable differences in the correlation properties of unobscured and obscured AGN have been reported.  For our simplified calculation we fixed the equivalent width of the narrow iron line at the value of $300$ eV, which correctly reproduces contribution of line photons to the CXB, $\sim 5-10\%$ at a few keV (see Section~\ref{sec2}), obtained in more  elaborate spectral synthesis calculations \citep{1999NewA....4...45G}. We investigate the dependence of our results on the assumed iron line strength in Section~\ref{sec3}, where we repeat our calculations for lower and higher values of the line equivalent width.
The dispersion width of the line was fixed at $45$ eV FWHM, in agreement with the Chandra grating observations of a large sample of local AGN \citep{2010ApJS..187..581S,2011ApJ...738..147S}. In order to investigate dependence of our results on the line width and to allow for a possible uncertainty in its measurements we also ran some of our calculations for a three times narrower line, i.e., $15$ eV FWHM (see, e.g., Fig.~\ref{fig5}).

In what follows, we use the notation where all the quantities that correspond to counting particles are written with a tilde on top, e.g., the luminosity is written as $\widetilde{L}$ and measured in units of s$^{-1}$, etc.

To calculate the two-point functions of the CXB we use the halo model approach (HM) (see \citet{2002PhR...372....1C} for an extensive review). According to HM the pairs of points are separated into two classes: (i) both points inside the same DM halo (one-halo term), which describes two-point function on small scales, (ii) points in separate DM halos (two-halo term), providing a large-scale two-point correlator. By making the assumption that AGN always reside at the centers of DM halos (which might be quite a good assumption, e.g., \citet{2011ApJ...741...15S}) they provide simply a constant one-halo term, which can be written as
\begin{eqnarray}\label{eq4}
C_{\ell}^{(ij),\,1h}&=&\frac{1}{{\cal \widetilde{F}}^{(i)}{\cal \widetilde{F}}^{(j)}}\int\int{\rm d}z\frac{{\rm d}V_c}{{\rm d}z}(z){\rm d}L\frac{{\rm d}n}{{\rm d}L}(L,z)\nonumber\\
&\times&\widetilde{F}^{(i)}(L,z)\widetilde{F}^{(j)}(L,z)\,.
\end{eqnarray} 
Here the superscripts $i$ and $j$ are used to denote two observational energy intervals. In general, Eq.~(\ref{eq4}) gives us one-halo coss-spectra between energy bins $i$ and $j$, while the case $i=j$ provides auto-spectra. $\frac{{\rm d}V_c}{{\rm d}z}$ is the comoving volume element per steradian and $\widetilde{F}^{(i)}$ is the photon flux received from a single AGN (redshift $z$ and $2-10$ keV luminosity $L$) in the observational energy range $E^{(i)}_{\min}-E^{(i)}_{\max}$, i.e.,
\begin{equation}
\widetilde{F}^{(i)}(L,z)=\frac{1+z}{4\pi d_L^2(z)}\int\limits^{(1+z)E^{(i)}_{\max}}_{(1+z)E^{(i)}_{\min}}\widetilde{L}_E(L){\rm d}E\,,
\end{equation}
where $d_L$ is luminosity distance and $\widetilde{L}_E$ the AGN spectral template (as described above), which is normalized such that
\begin{equation}
L=\int\limits^{10\, {\rm keV}}_{2\, {\rm keV}}E\widetilde{L}_E{\rm d}E\,.
\end{equation}
In Eq.~(\ref{eq4}) ${\cal \widetilde{F}}^{(i)}$ is the total photon flux received in the energy interval $i$
\begin{equation}\label{eq7}
{\cal \widetilde{F}}^{(i)}=\int\int{\rm d}z\frac{{\rm d}V_c}{{\rm d}z}(z){\rm d}L\frac{{\rm d}n}{{\rm d}L}(L,z)\widetilde{F}^{(i)}(L,z)\,.
\end{equation}

The corresponding two-halo spectra can be given as
\begin{equation}\label{eq8}
C_{\ell}^{(ij),\,2h}=\frac{2}{\pi}\int W^{(i)}_{\ell}(k)W^{(j)}_{\ell}(k)P^{\rm lin}(k)k^2{\rm d}k\,,
\end{equation}
where $P^{\rm lin}(k)$ is the 3D linear power spectrum at $z=0$ and the projection kernels (see, e.g., \citet{2001ApJ...555..547H,2007MNRAS.378..852P} for details) are given as~\footnote{In principle, one could also include the effects of redshift-space distortions here, following \citet{2007MNRAS.378..852P}. However, this would only lead to a noticeable difference on scales larger than the scales where most of our signal tends to arise. Also, to have enough photon statistics available, we cannot make the observational energy range too narrow, so even photons emitted from the $6.4$ keV line originate in a relatively broad redshift shell, which makes redshift distortions quite negligible in practice. Due to these reasons we have chosen not to include redshift-space distortions in our calculations.}
\begin{equation}\label{eq9}
W^{(i)}_{\ell}(k)=\int j_{\ell}(kr)g(r)b_{\rm eff}^{(i)}(r)f^{(i)}(r){\rm d}r\,.
\end{equation}
Here the integral is over comoving distance $r$, $j_{\ell}$ is the spherical Bessel function, $g$ the linear growth factor, $b_{\rm eff}^{(i)}$ the effective clustering bias in energy bin $i$, which can be given as
\begin{equation}
b_{\rm eff}^{(i)}(z)=\frac{\int {\rm d}L\frac{{\rm d}n}{{\rm d}L}(L,z)b\left[M(L),z\right]\widetilde{F}^{(i)}(L,z)}{\int {\rm d}L\frac{{\rm d}n}{{\rm d}L}(L,z)\widetilde{F}^{(i)}(L,z)}\,.
\end{equation}
To calculate $b(M,z)$ we use the analytical model of \citet{2001MNRAS.323....1S}.
$f^{(i)}$ in Eq.~(\ref{eq9}) is the radial selection function, i.e., the probability distribution function for the emission redshifts of photons. This quantity can be expressed as
\begin{equation}
f^{(i)}(z)=\frac{\int {\rm d}L\frac{{\rm d}V_c}{{\rm d}z}(z)\frac{{\rm d}n}{{\rm d}L}(L,z)\widetilde{F}^{(i)}(L,z)}{{\cal \widetilde{F}}^{(i)}}\,,
\end{equation}
where the normalization factor ${\cal \widetilde{F}}^{(i)}$, i.e. the total photon flux in energy bin $i$, is given by Eq.~(\ref{eq7}). In all the equations above, where the integration over AGN luminosities is performed, we have assumed the lower integration bound to be $10^{41}$ erg/s, and the higher bound is taken to be equal to the luminosity, which corresponds to the limiting flux $F_{\rm lim}$ above which sources are removed. 

The radial selection functions for Models I and II are shown in the upper panel of Fig.~\ref{fig1}. There we have assumed energy bins $3.0-3.2$ keV and $3.3-3.5$ keV~\footnote{To be more precise, to ensure that the continuum contributes equally to both energy bins, the upper energy in bin 2, $E^{(2)}_{\max}$, is calculated as (valid if $\Gamma=2$) $E^{(2)}_{\max}=(1/E^{(2)}_{\min}-1/E^{(1)}_{\min}+1/E^{(1)}_{\max})^{-1}\simeq0.3544$ keV. However, when the energy bin gets narrow, this difference does not matter much, and $E^{(2)}_{\max}=3.5$ keV gives practically equivalent results for the final power spectra.}, and have taken the flux above which sources are removed, $F_{\rm lim}$, equal to $10^{-13}$ erg/s/cm$^2$. Thus, the angular power spectra according to the HM are calculated as
\begin{equation}
C_{\ell}^{(ij)}=C_{\ell}^{(ij),\,2h}+C_{\ell}^{(ij),\,1h}\,,
\end{equation}
where the two- and one-halo terms are given by Eqs.~(\ref{eq8}) and (\ref{eq4}), respectively.

\begin{figure}
\centering
\includegraphics[width=\plotwd]{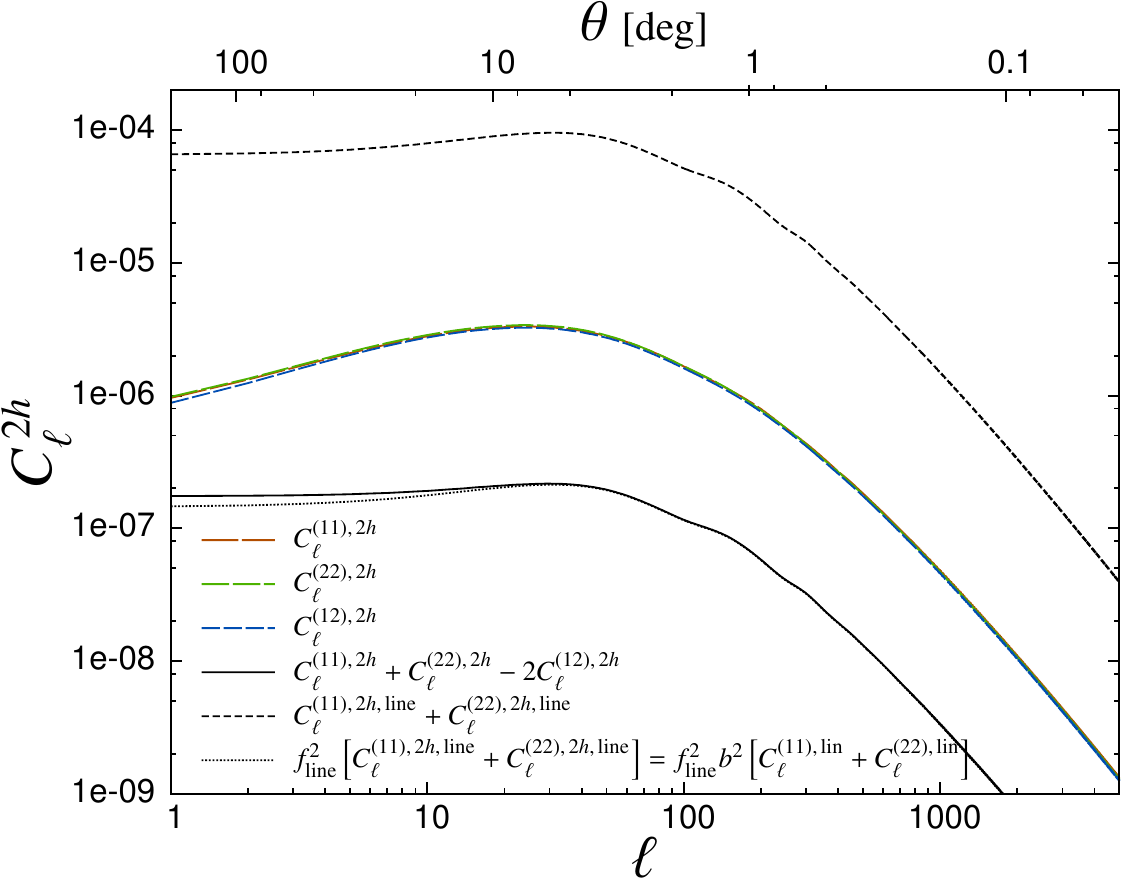}
\caption{Two-halo terms of angular auto and cross power spectra for Model I assuming two observational energy ranges: (1) $3.0-3.2$ keV and (2) $3.3-3.5$ keV and fixing $F_{\rm lim}$ to $10^{-13}$ erg/s/cm$^2$, as in Fig.~\ref{fig1}. The three dashed lines with intermediate amplitude represent auto-spectra in energy bin 1 ($C_{\ell}^{(11),2h}$) and bin 2 ($C_{\ell}^{(22),2h}$) along with the corresponding cross-spectrum ($C_{\ell}^{(12),2h}$) (all three lines are on top of each other, with $C_{\ell}^{(12),2h}$ being slightly lower at small $\ell$). The solid line shows the signal after removal of the continuum part, i.e., $C_{\ell}^{(11)}+C_{\ell}^{(22)}-2C_{\ell}^{(12)}$. The short-dashed line with the highest amplitude corresponds to the sum of autocorrelations of bin 1 and 2 in case there is only signal from $6.4$ keV line and no continuum contribution. The same curve multiplied by $f_{\rm line}^2$ (the square of fraction of photons from the line) is plotted as a dotted line, which is also equivalent to the corresponding sum of linear spectra multiplied by $b^2f_{\rm line}^2$, where $b$ is the linear clustering bias parameter (in current case $b\simeq 2.5$ and $f_{\rm line} \simeq 0.047$).}
\label{fig2}
\end{figure}

\subsection{Line signal. Subtraction of the dominant continuum contribution}
To extract the line signal from the dominant continuum contribution, we calculate the following quantity
\begin{equation}\label{eq13}
C_{\ell}=C_{\ell}^{(11)}+C_{\ell}^{(22)}-2C_{\ell}^{(12)}\,;
\end{equation}
i.e., from the sum of the auto-spectra of two nearby energy bins we subtract their double cross-spectrum. It is clear that this procedure only removes the slowly changing component (continuum) since then
\begin{equation}\label{eq14}
C_{\ell,\, {\rm cont}}^{(11)}\simeq C_{\ell,\, {\rm cont}}^{(22)}\simeq C_{\ell,\, {\rm cont}}^{(12)}\,,
\end{equation}
and we are basically left with only the line signal. Of course, for this procedure to work accurately enough, i.e., Eq.~(\ref{eq14}) to be valid, one has to adjust energy ranges $E^{(1)}_{\min}-E^{(1)}_{\max}$ and $E^{(2)}_{\min}-E^{(2)}_{\max}$ so that continuum gives equal contribution (in terms of detected number of photons) to both. This seems to demand some prior information on the effective continuum shape, which one need not have available. However, in practice, when one looks at narrow enough energy ranges, where the continuum changes only slightly in contrast to the rapidly varying line component, this scheme should perform reasonably well, even when we have no detailed information on the smooth continuum part.

In Fig.~\ref{fig2} we show $C_{\ell}^{(11)}$, $C_{\ell}^{(22)}$ and $C_{\ell}^{(12)}$ along with the resulting signal calculated from Eq.~(\ref{eq13}). Here we have plotted two-halo components only. Also, as the curve with the largest amplitude, we show the signal in case there is only line contribution without continuum. The observing energy ranges are chosen the same way as in Fig.~\ref{fig1} to ensure that line contributions have no overlap in $z$, and thus the cross-spectrum $C_{\ell}^{(12),\,{\rm line}}\simeq 0$. If the mean photon flux from the line in the energy range $E^{(1)}_{\min}-E^{(1)}_{\max}$ ($E^{(2)}_{\min}-E^{(2)}_{\max}$) is the fraction $f^{(1)}_{\rm line}$ ($f^{(2)}_{\rm line}$) of the total flux in that energy interval, the signal $C_{\ell}$ is related to the ``line only'' signal as
\begin{eqnarray}\label{eq15}
C_{\ell}&\simeq& \left(f^{(1)}_{\rm line}\right)^2C_{\ell}^{(11),\,{\rm line}} + \left(f^{(2)}_{\rm line}\right)^2 C_{\ell}^{(22),\,{\rm line}}\nonumber\\
&\simeq& f^2_{\rm line}\left[C_{\ell}^{(11),\,{\rm line}} + C_{\ell}^{(22),\,{\rm line}}\right]\,,
\end{eqnarray}
because in practice $(f^{(1)}_{\rm line}\simeq f^{(2)}_{\rm line})\equiv f_{\rm line}$.
This can also be given as
\begin{equation}
C_{\ell}\simeq f^2_{\rm line}b^2\left[C_{\ell}^{(11),\,{\rm lin}} + C_{\ell}^{(22),\,{\rm lin}}\right]\,,
\end{equation}
where $C_{\ell}^{(11),\,{\rm lin}}$ and $C_{\ell}^{(22),\,{\rm lin}}$ are the linear density fluctuation spectra in case the radial selection corresponds to the ``line only'' selections of Fig.~\ref{fig1}, and $b$ is the clustering bias parameter (in this particular case $b\simeq 2.5$ and $f_{\rm line} \simeq 0.047$).

The quantity on the right-hand side of Eq.~(\ref{eq15}), i.e., the line signal we wish to recover, is plotted as a dotted line in Fig.~\ref{fig2}, demonstrating that our approximate scheme of Eq.~(\ref{eq13}) for removing the continuum component (shown with a solid line) works quite well. The small deviations at low $\ell$ are due to residual line-continuum cross terms, which are missing in the ``line only'' case.  

Thus, the simplest quantity one can hope to obtain from measuring of $C_{\ell}$ is the signal amplitude $A=bf_{\rm line}$, if the linear density fluctuation spectrum is known. The analysis of how well one can determine the amplitude $A$ is presented in Section~\ref{sec3} below.

\begin{figure}
\centering
\includegraphics[width=\plotwd]{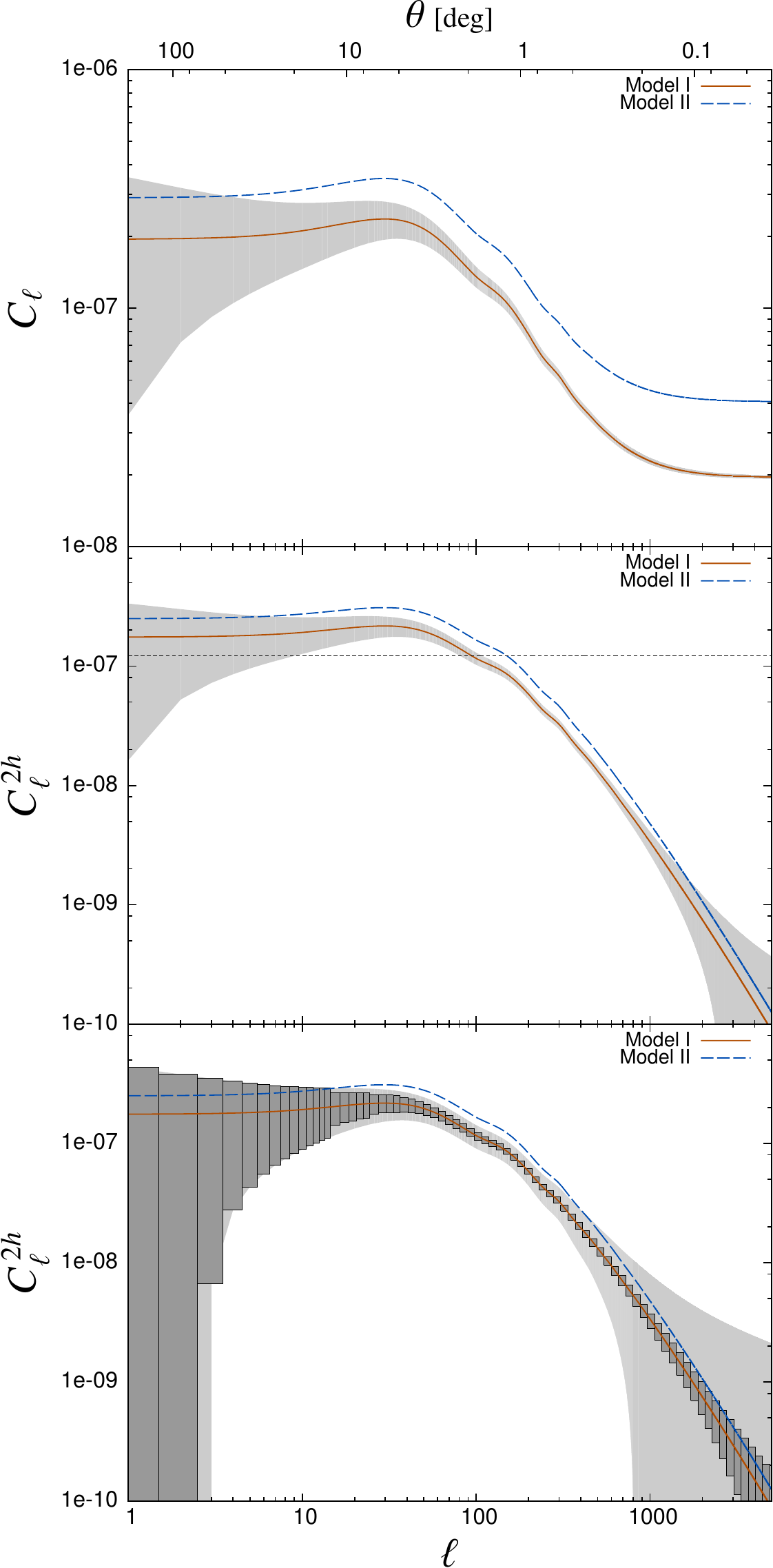}
\caption{Angular power spectra for the extracted line signal assuming energy bins $3.0-3.2$ keV and $3.3-3.5$ keV along with $F_{\rm lim}=10^{-13}$ erg/s/cm$^2$. For clarity the errors are only displayed for Model I. To calculate the level of photon noise we have assumed that flux $10^{-13}$ erg/s/cm$^2$ corresponds to $500$ counts. {\bf Upper panel:} Angular spectra including two- and one-halo terms. The gray shaded area shows $1\sigma$ uncertainty region as calculated from Eq.~(\ref{eq18}). Here we have assumed perfect photon sampling, i.e., the photon noise is taken to be zero. {\bf Middle panel:} Angular power spectra with one-halo term subtracted. The gray band shows error corridor as calculated from Eq.~(\ref{eq19}). The level of photon noise, which is not yet included in calculation of errors, is shown by a short-dashed line. {\bf Lower panel:} The same as middle panel but with the effect of photon noise included (Eq.~(\ref{eq21})). The light gray band and dark gray histogram present errors without and with binning, respectively.}
\label{fig3}
\end{figure}

\subsection{Error estimates}

Assuming that the fluctuation fields follow Gaussian statistics, which is a valid assumption on large enough scales~\footnote{For ``large enough'' we mean scales that are described well by the linear theory. It turns out that information on signal amplitude $A$ arises mostly from modes $\ell \sim 100$, which assuming the typical redshift range of the dominant AGN activity ($z\sim 1$), correspond to comoving scales that are indeed well within the linear regime.}, the errors on spectra $C_{\ell}^{(ij)}$ can be written as \citep{1995PhRvD..52.4307K,1996PhRvD..54.1332J}
\begin{equation}\label{eq17}
\delta C_{\ell}^{(ij)}=\sqrt{\frac{2}{(2\ell+1)f_{\rm sky}}}\cdot C_{\ell}^{(ij)}\,,
\end{equation}
where $f_{\rm sky}$ is the fraction of sky covered by the survey. Throughout this study we take $f_{\rm sky}=1$. The results can be easily rescaled to a more realistic value of $f_{\rm sky}\approx 0.83$ \citep{1982ApJ...255..111W,2006A&A...452..169R}, corresponding to the extragalactic sky $|b|>10\degr$.~\footnote{Unlike the CMB studies, where the Galactic foreground emission is bright in a broad range of Galactic latitudes, the bright Galactic ridge X-ray emission is more concentrated toward the Galactic plane.} 

Since the signals in nearby energy bins are highly correlated, being dominated by the continuum contribution, we obtain for $\delta C_{\ell}$
\begin{equation}\label{eq18}
\delta C_{\ell} = \delta C_{\ell}^{(11)}+\delta C_{\ell}^{(22)}-2\delta C_{\ell}^{(12)}=\sqrt{\frac{2}{(2\ell+1)f_{\rm sky}}}\cdot C_{\ell}\,.
\end{equation}
The last result is similarly obtained if one realizes that $C_{\ell}$ of Eq.~(\ref{eq13}) is in reality a power spectrum of the difference field (fluctuation field in the first energy range minus the field in the second energy range). If both fluctuation fields can be approximated as Gaussian, so can be any linear combination of these, then Eq.~(\ref{eq18}) follows immediately.

\begin{figure}
\centering
\includegraphics[width=\plotwd]{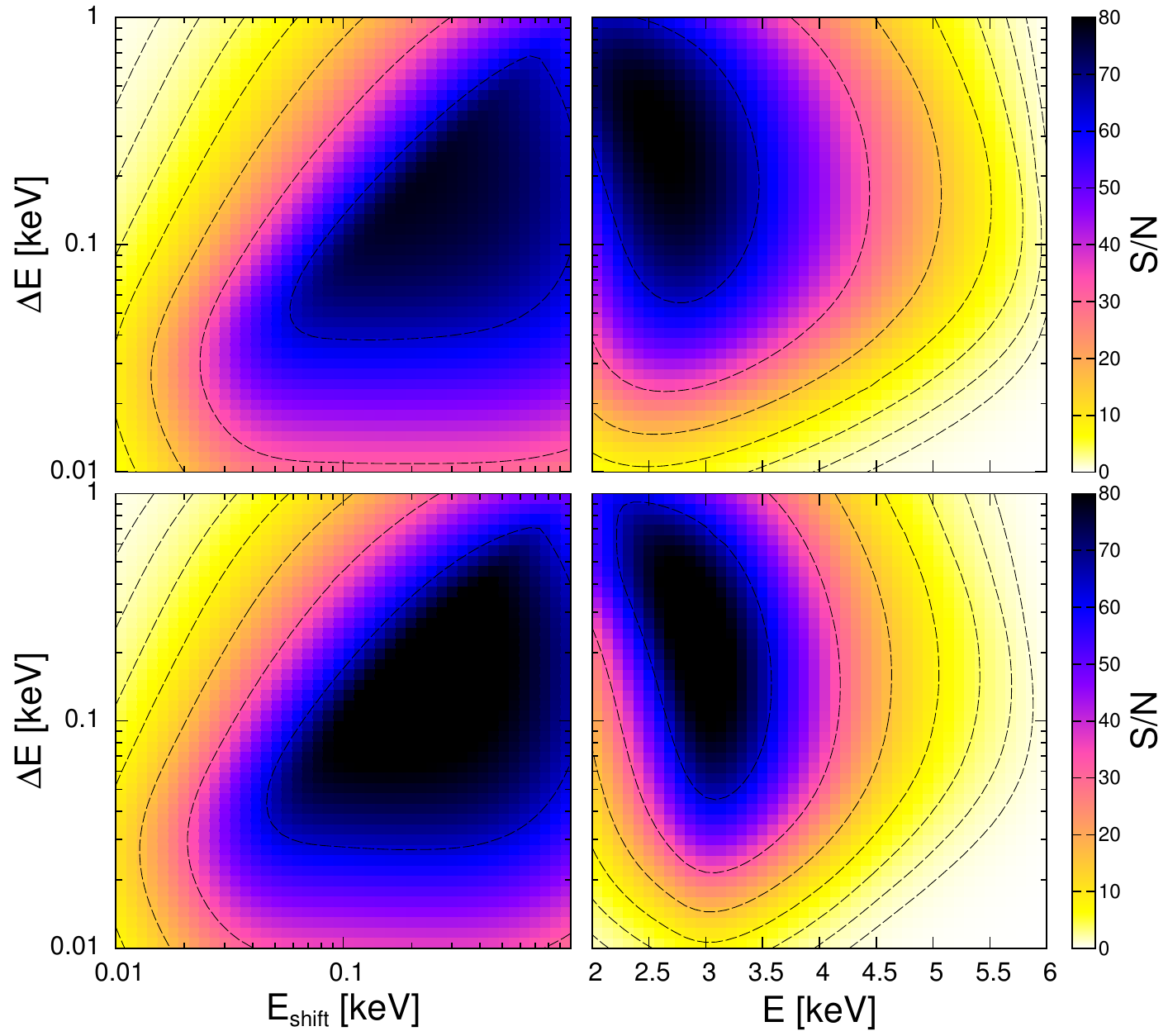}
\caption{Signal-to-noise ratio according to Eq.~(\ref{eq22}) with $\ell_{\max}=500$. $F_{\rm lim}=10^{-13}$ erg/s/cm$^2$ along with $500$ counts at this flux is assumed. The upper and lower panels correspond to Models I and II, respectively. In the left-hand panels we show $S/N$ as a function of $\Delta E$ and $E_{\rm shift}$ while keeping $E$ fixed to $3.0$ keV. The right-hand panels display $S/N$ as a function of $E$ and $\Delta E$ by keeping $E_{\rm shift}$ equal to $\Delta E$. The dashed $S/N$ contours correspond to values $1$, $2$, $4$, $8$, $16$, and $32$.}
\label{fig4}
\end{figure}

In the upper panel of Fig.~\ref{fig3} we show $C_{\ell}$ for Models I and II. The observed energy ranges are the same as in Fig.~\ref{fig1}. For clarity, the error range, which is shown by the shaded area, is given only for Model I. In the middle panel we present only the two-halo terms, i.e., the one-halo contributions are subtracted. As the amplitude of the one-halo term can be determined with very good accuracy, the errors in this case are also approximated well by
\begin{equation}\label{eq19}
\delta C_{\ell}^{(2h)}\simeq \delta C_{\ell} = \sqrt{\frac{2}{(2\ell+1)f_{\rm sky}}}\cdot \left(C_{\ell}^{2h}+C_{\ell}^{1h}\right)\,.
\end{equation}
In this panel we have also shown the photon noise level, which is taken to be
\begin{equation}
{\rm SN}=\sqrt{\left(\frac{1}{N^{(1)}}\right)^2+\left(\frac{1}{N^{(2)}}\right)^2}\,,
\label{eq:phot_noise}
\end{equation}
where $N^{(1)}$ and $N^{(2)}$ are the registered counts per steradian in the first and second energy bins. Here we have assumed that the photon noise can be taken to be independent in those two non-overlapping energy ranges, and that the cross-spectrum is free of photon noise.

In the last panel of Fig.~\ref{fig3} we show the errors once the photon noise is included; i.e.,
\begin{equation}\label{eq21}
\delta C_{\ell}^{(2h)}\simeq \sqrt{\frac{2}{(2\ell+1)f_{\rm sky}}}\cdot \left(C_{\ell}^{2h}+C_{\ell}^{1h}+{\rm SN}\right)\,.
\end{equation}
The light gray shaded area corresponds to the unbinned case, while the dark gray histogram represents the binning with $\Delta \ell=0.1\ell$. 

As one can see from Fig.~\ref{fig3}, predictions of the two models agree within $\sim 30\%$

\section{Results}\label{sec3}
\subsection{Expectations for the signal strength}
In this section we investigate how well the signal amplitude $A=bf_{\rm line}$ could be measured. Focusing only on this single parameter, we can easily write for it the signal-to-noise ratio as
\begin{equation}\label{eq22}
S/N\equiv \frac{A}{\delta A}=\sqrt{\sum\limits_{\ell}\left(\frac{C_{\ell}^{(2h)}}{\delta C_{\ell}^{(2h)}}\right)^2}\,.
\end{equation}
The summation over the multipole number $\ell$ is truncated at $\ell_{\max}$, up to which one can still assume the validity of the linear theory along with the Gaussianity assumption. In 3D the corresponding comoving wavenumber is often taken to be $k_{\max}=0.2$ hMpc$^{-1}$. As the strongest signal is expected for the energies that sample the maximum of the radial selection function, which occurs at $z\sim 1$ (see Fig.~\ref{fig1}), we can write $\ell_{\max}=k_{\max}R_{z=1}\simeq 0.2\cdot 2400\simeq 500$ in this case, where $R_{z=1}$ is the comoving distance to $z=1$. In reality one should vary $\ell_{\max}$ depending on at what redshift most of the signal originates. However, as our interest is mostly concerned with energies (redshifts) where the signal-to-noise ratio is significantly high, and this turns out to be constrained to a rather narrow energy (redshift) range, we can safely keep $\ell_{\max}$ fixed to the above value. Also, it turns out that most of the sensitivity for measuring $A$ comes from the multipoles $\ell$, which are somewhat smaller than $\ell_{\max}$, so the case without upper $\ell$ cutoff in Eq.~(\ref{eq22}) only mildly improves the signal-to-noise ratio.

\begin{figure}
\centering
\includegraphics[width=\plotwd]{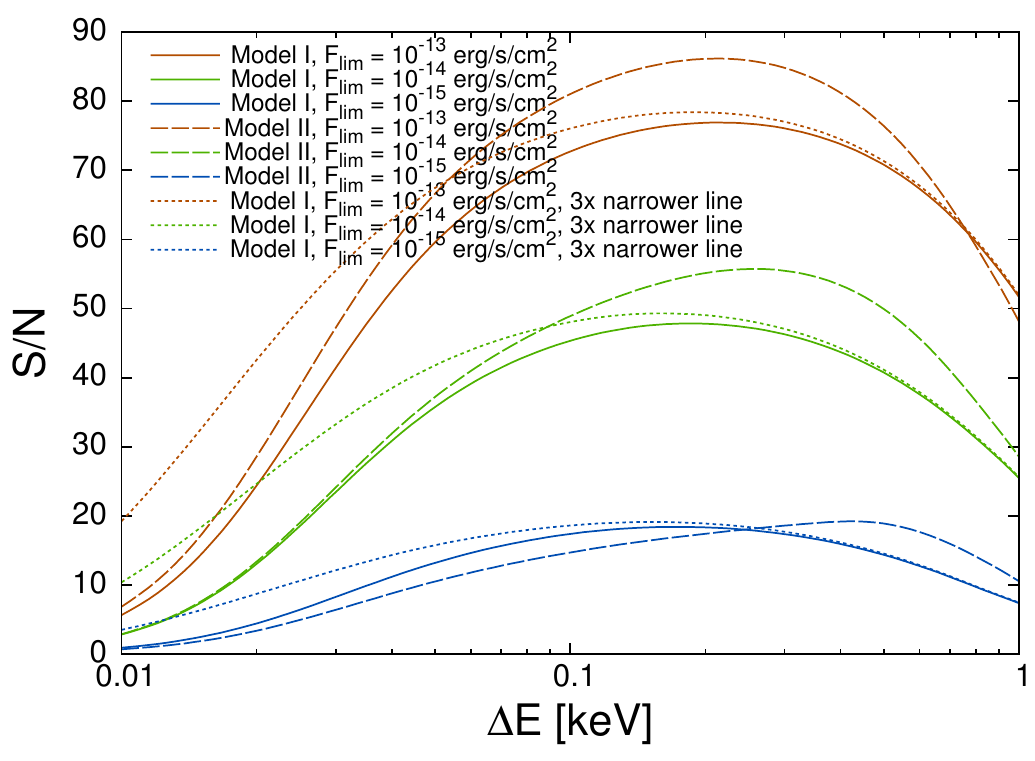}
\caption{Signal-to-noise for Models I (solid lines) and II (dashed lines) as a function of $\Delta E$ keeping $E$ fixed to $3.0$ keV and allowing $F_{\rm lim}$ to take values $10^{-13}$ (upper curves), $10^{-14}$ (middle curves), and $10^{-15}$ erg/s/cm$^2$ (lower curves), while keeping the flux to counts conversion factor the same, i.e., $500$ counts at $10^{-13}$ erg/s/cm$^2$. The dotted lines are for Model I with a three times narrower line, i.e., $15$ eV FWHM.}
\label{fig5}
\end{figure}

\begin{figure}
\centering
\includegraphics[width=\plotwd]{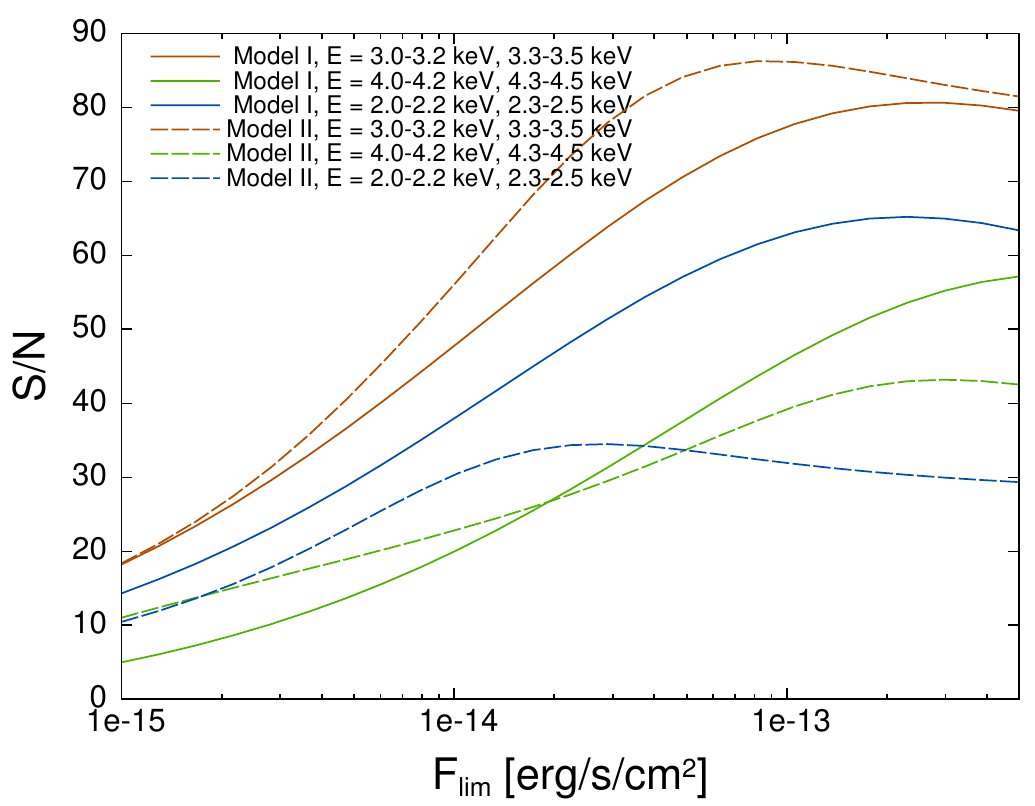}
\caption{Signal-to-noise for Models I and II (solid and dashed lines, respectively) as a function of $F_{\rm lim}$ assuming observational energy ranges as shown in the legend. The flux to photon count conversion factor is kept fixed to $500$ counts at $10^{-13}$ erg/s/cm$^2$.}
\label{fig6}
\end{figure}

\begin{figure}
\centering
\includegraphics[width=\plotwd]{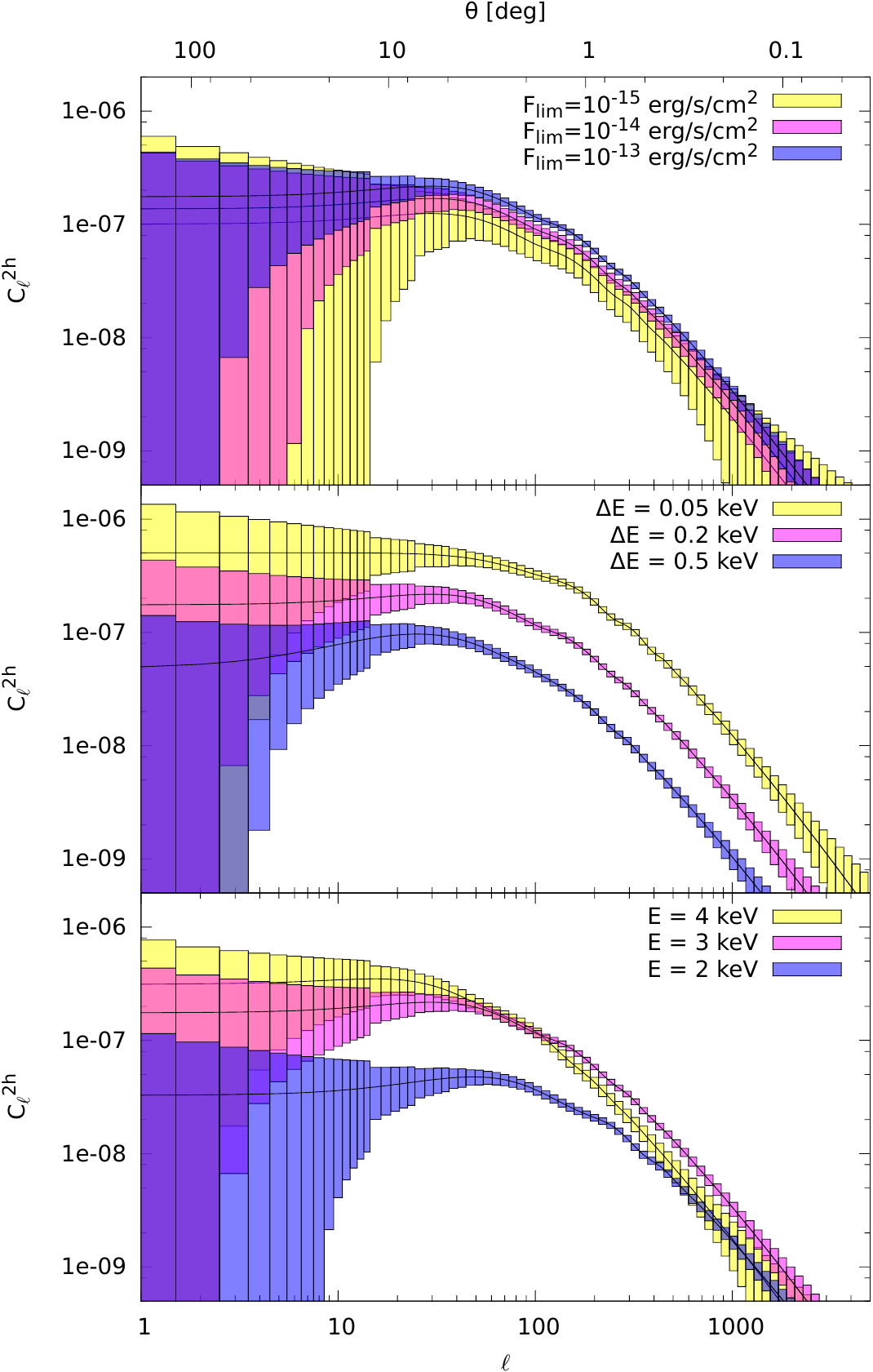}
\caption{The changes in the extracted line signal angular power spectra for Model I by allowing $F_{\rm lim}$ (upper panel), $\Delta E$ (middle panel), and $E$ (lower panel) to vary. In the upper panel $E=3.0$ keV, $\Delta E=0.2$ keV, in the middle panel $E=3.0$ keV, $F_{\rm lim}=10^{-13}$ erg/s/cm$^2$, and in the lower panel $\Delta E=0.2$ keV, $F_{\rm lim}=10^{-13}$ erg/s/cm$^2$. The flux-to-photon count conversion factor is again fixed to $500$ counts at $10^{-13}$ erg/s/cm$^2$.} 
\label{fig7}
\end{figure}

In the following we study signal-to-noise as a function of width and location of the energy bins. We also investigate the dependence on the limiting flux $F_{\rm lim}$ above which sources are removed. The conversion factor between photon counts and source flux is set to $500$ counts at $10^{-13}$ erg/s/cm$^2$ ($2-10$ keV band), as before. This level of photon statistics is achievable with, e.g., $\sim 25$ ksec XMM-Newton exposure using PN+2MOS detectors.

We denote the beginning of the first energy bin $E^{(1)}_{\min}$ by $E$, the width $E^{(1)}_{\max}-E^{(1)}_{\min}$ is represented by $\Delta E$, and the distance between the bins $E^{(2)}_{\min}-E^{(1)}_{\min}$ is written as $E_{\rm shift}$. The upper energy of the second bin $E^{(2)}_{\max}$ is calculated as $E^{(2)}_{\max}=(1/E^{(2)}_{\min}-1/E^{(1)}_{\min}+1/E^{(1)}_{\max})^{-1}$ (valid if $\Gamma=2$), which guarantees that the continuum contributes equally to both energy bins. Thus, including the limiting flux $F_{\rm lim}$, we have in total four parameters which we choose to vary to study the effect on signal-to-noise ratio.

In the left-hand panels of Fig.~\ref{fig4} we have fixed $F_{\rm lim}=10^{-13}$ erg/s/cm$^2$, $E=3.0$ keV, and allowed $\Delta E$ and $E_{\rm shift}$ to vary in logarithmic steps between $10^{-2}$ and $1$ keV. The upper and lower panels correspond to Models I and II, respectively. As one might have expected, we see that signal-to-noise ratio achieves highest values if one chooses $E_{\rm shift}\simeq \Delta E$, since this guarantees that there is not much overlap between the redshift ranges where the line signal of both energy bins originates. Above the diagonal line $E_{\rm shift}=\Delta E$, the line contributions start to overlap, which leads to a fast drop in signal-to-noise. Also, the signal-to-noise ratio starts to decrease rapidly (mostly due to limited photon statistics) in case one chooses too small $\Delta E$, smaller than the assumed line width. As with $E=3.0$ keV we are probing redshifts $z\simeq 1$, $45$ eV FWHM corresponds to $0.045/(1+z)\simeq 0.023$ keV for the $z=0$ observer.

In the right-hand panels of Fig.~\ref{fig4} we have fixed $E_{\rm shift}=\Delta E$, and allowed $E$ and $\Delta E$ to vary. We see that the highest signal-to-noise is achieved for the energies $E=2-4$ keV, which corresponds to the redshift range where the radial selection shown in Fig.~\ref{fig1} has considerable amplitude. We can also see that, with only a mild dependence on observational energy, the highest achievable signal-to-noise ratio corresponds to $\Delta E\sim 0.2$ keV, with quite a broad maximum around that value. This is best seen in Fig.~\ref{fig5}, where we have shown cuts through the right-hand panels of Fig.~\ref{fig4} at fixed $E=3.0$ keV. Here, in addition to $F_{\rm lim}=10^{-13}$ erg/s/cm$^2$, we have also shown the cases with $F_{\rm lim}=10^{-14}$ and $10^{-15}$.

In Fig.~\ref{fig6} we show how signal-to-noise ratio varies as a function of $F_{\rm lim}$, by fixing $\Delta E=0.2$ keV and $E_{\rm shift}=0.3$ keV, for the observing energies
 $E=2,3,$ and $4$ keV. As noted before, the optimal signal-to-noise is achieved if $E\sim 3$ keV, and it drops appreciably if one goes $\sim 1$ keV above or below that value. Similarly, one sees that optimal values for $F_{\rm lim}$, above which to remove bright sources, are around $10^{-13}$ erg/s/cm$^2$, while signal-to-noise drops by a factor $\sim 2-4$ if one reduces $F_{\rm lim}$ down to $10^{-15}$ erg/s/cm$^2$.

The underlying changes in the power spectra, along with error bars, for Model I are shown in Fig.~\ref{fig7}. Here in the upper panel $E=3.0$ keV, $\delta E=E_{\rm shift}$, and we have varied $F_{\rm lim}$ from $10^{-15}$ up to $10^{-13}$ erg/s/cm$^2$. By reducing the value of $F_{\rm lim}$, we see how the error bars get progressively larger, and the signal amplitude drops due to the removal of brighter, and thus (in this model) more biased, sources.

In the middle panel of Fig.~\ref{fig7} the effect of varying $\Delta E (=E_{\rm shift})$, while keeping $E=3.0$ keV and $F_{\rm lim}=10^{-13}$, is shown. Again, the error bars are smallest for the case $\Delta E=0.2$ keV, and increase in the other two cases. The amplitude of the signal keeps on increasing, as one would expect, since narrower $\Delta E$ corresponds to the narrower radial selection function, hence less smearing along the line of sight.

In the last panel we have varied the observing energies ($E=2,3,4$ keV), while fixing $\Delta E=0.2$ and $F_{\rm lim}=10^{-13}$, and so effectively probe cosmic structure at different redshifts, which explains the largest variation in spectral shapes seen among the panels of Fig.~\ref{fig7}. Again, the tightest error bars correspond to $E=3$ keV case.  

\subsection{Prospects for current and future X-ray instruments}

In this section we investigate the potential of the current, near-term, and proposed future X-ray instruments to  detect  the $6.4$ keV line tomographic signal. We consider all three major currently operating  X-ray observatories -- Chandra, XMM-Newton and Suzaku~\footnote{\url{http://www.astro.isas.ac.jp/suzaku/}}, two  missions planned for a launch in the next few years -- ASTRO-H~\footnote{\url{http://astro-h.isas.jaxa.jp/}} and eROSITA, and several proposed   mission concepts -- ATHENA, WFXT, and SMART-X~\footnote{\url{http://hea-www.cfa.harvard.edu/SMARTX/}}. We also include LOFT, due to its large grasp, although it does not carry an imaging instrument.   Finally, we consider a hypothetical future mission with the effective area of 10 m$^2$ and the field of view of 1 deg$^2$. 

\begin{figure}
\centering
\includegraphics[width=\plotwd]{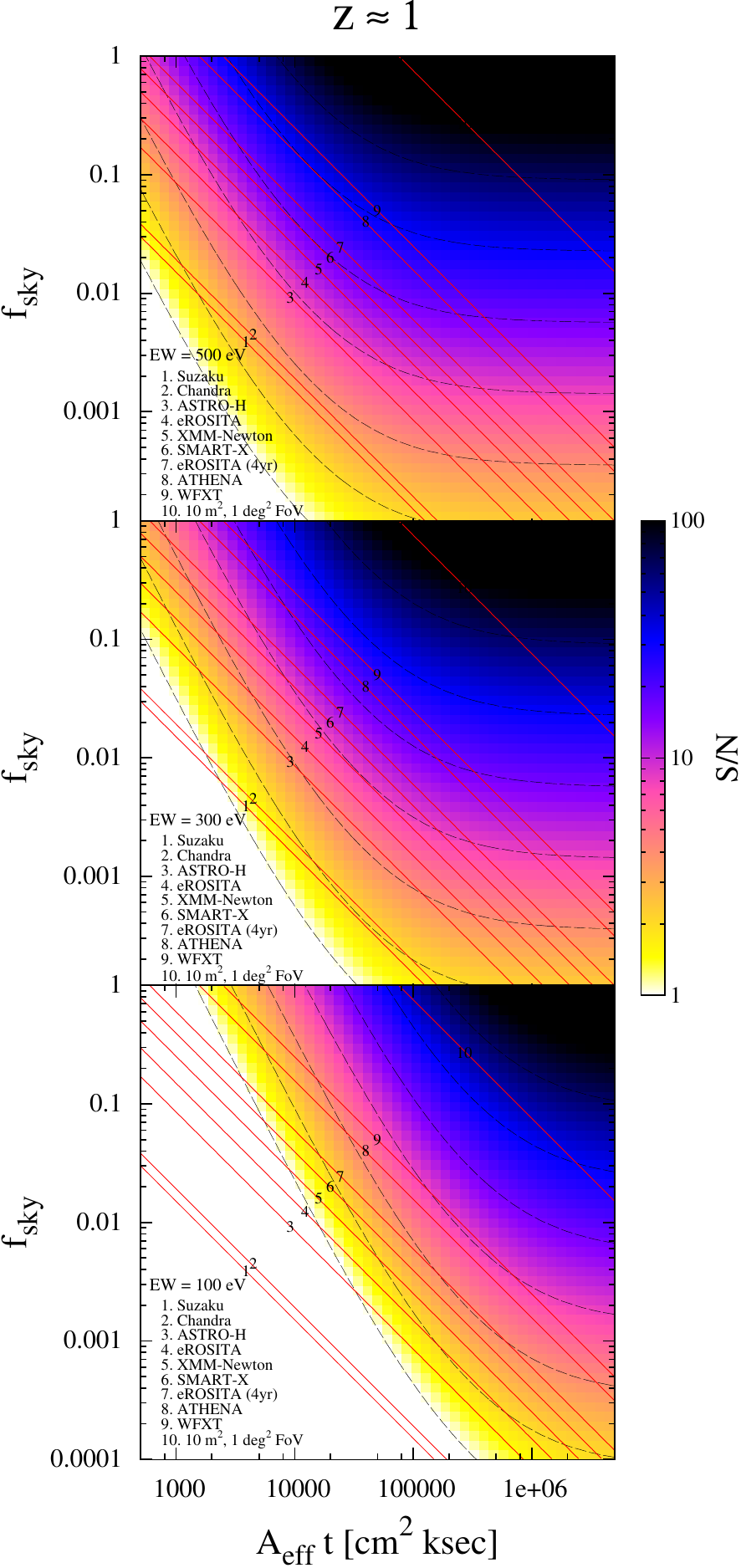}
\caption{Signal-to-noise ratio of the tomographic signal as a function of the  fraction of sky $f_{sky}$ covered in the survey and $A_{\rm eff}\times t$ -- a product of the effective area $A_{\rm eff}$, at the energy where the tomographic signal is extracted, and the time $t$ spent per field. The dashed $S/N$ contours correspond to values $1$, $2$, $4$, $8$, $16$, $32$, and $64$. The straight lines show the locus of the points, which can be achieved in the course of a 1-year survey (4 years for eROSITA) by various currently existing, near-term, and possible future X-ray instruments. The panels show results for different values of the  $6.4$ keV line equivalent width:  $500$ eV (top panel), $300$ eV (middle) and $100$ eV (bottom). Observational energy bands $3.0-3.2$ keV and $3.3-3.5$ keV, corresponding to $z\simeq 1$, are assumed.}
\label{fig8}
\end{figure}

So far we have assumed  perfect energy resolution of the telescope, whereas the energy response of the majority of  imaging  instruments above (not considering micro-calorimeters) have a width of $\sim 130$ eV (FWHM) at $6$ keV.  This typically translates to $\sim 90$ eV at $3$ keV, where most of the signal is expected to arise.  Assuming that instrumental energy response is approximated well by a Gaussian, the instrumental energy resolution  can be accounted for by substituting the intrinsic line width with the effective width (convolving a Gaussian with another Gaussian produces a Gaussian):
\begin{equation}\label{newsec_eq1}
\sigma_{\rm eff}=\sqrt{\sigma_{\rm line}^2+(1+z)^2\sigma_{\rm instr}^2}\,.
\end{equation}
The instrumental energy resolution is included in calculations of this section using the above formula.
 
As it follows from the discussion of the noise components (Eqs.~(\ref{eq19}) and (\ref{eq:phot_noise})), the signal-to-noise ratio $S/N$ is determined by the fraction of sky $f_{sky}$ covered in the survey and the average number of counts registered per FOV.
To characterize the latter we use the  quantity $A_{\rm eff}\times t$, a product of the effective area $A_{\rm eff}$ at the energy where the tomographic signal is extracted and the time $t$ spent per field. For a flat energy response, $A_{\rm eff}\times t \, [{\rm cm^2\,ksec}]\simeq 65\times N_{\rm counts}$, where $N_{\rm counts}$ is the number of counts received in the 2--10 keV band from a $10^{-13}$ erg/s/cm$^2$ source with photon index $\Gamma=2$.  

The result of the signal-to-noise calculation for Model I is shown in Fig.~\ref{fig8} in  the form of a two-dimensional map, as a function of covered sky fraction $f_{\rm sky}$ and $A_{\rm eff}\times t$. 
In computing the photon counting noise we used the observed spectrum of the CXB and assumed flat response in the energy range of interest ($3.0-3.5$ keV)  to calculate $N_1$ and $N_2$ in Eq.~(\ref{eq:phot_noise}). The AGN discreteness term was computed self-consistently via Eq.~(\ref{eq19}). We note that the latter is somewhat  overestimated in our calculations since the $2-10$ keV band LF does not include the full contribution from Compton thick objects.

The middle panel in  Fig.~\ref{fig8} represents our default case corresponding to the $6.4$ keV line equivalent width of $300$ eV. To account for the large uncertainty in effective (population averaged) value of $6.4$ keV line strength we also show the results for $EW=500$ eV (upper panel) and $EW=100$ eV (lower panel). The choice of the energy bins ($3.0-3.2$ and $3.3-3.5$ keV) is tuned for the extraction of the tomographic signal at the redshift $z\simeq 1$, where it is strongest. We therefore take  $z\simeq 1$ in Eq.~(\ref{newsec_eq1}). The corresponding effective line width $\sim 150-200$ eV FWHM is rather large. However, as the energy bin widths are also large, $\Delta E\sim 0.2$ keV, and the signal strength has a rather broad maximum around this value (Fig.~\ref{fig5}), the additional signal smearing due to instrumental energy resolution turns out to have only mild effect on the signal-to-noise ratio.

In Fig.~\ref{fig8}, the signal-to-noise ratio increases as $\sqrt{f_{\rm sky}}$, and for low enough values of $A_{\rm eff}\times t$, i.e. when the dominant noise component is photon noise, it is proportional to $A_{\rm eff}\times t$. However, at large $A_{\rm eff}\times t$, the increase in signal-to-noise saturates because of the  presence of the irreducible noise component due to the discrete nature of AGN.  Correspondingly, the signal-to-noise iso-contours flatten out in the right-hand side of Fig.~\ref{fig8}.

To explore the suitability of different instruments for measuring the tomographic signal, we determine  the boundaries of the regions on the $f_{\rm sky}$--$A_{\rm eff}\times t$ plane, which can be achieved by these instruments. These boundaries are defined by the relation
\begin{equation}\label{newsec_eq2}
f_{\rm sky}=f_{\rm FoV}\frac{T}{t}=\frac{{\rm grasp}\times T}{A_{\rm eff}\times t}\,,
\end{equation}
where $T$  is the total time  spent for the survey, $t$ the time spent per one pointing, and $f_{\rm FoV}$ the fraction of sky covered by the field of view of the instrument. The key instrumental parameter determining the strength of the tomographic signal is the grasp,  $A_{\rm eff}\times f_{\rm FoV}$. The second parameter is, naturally,  the duration of the survey $T$. In applying the Eq.~(\ref{newsec_eq2}) to different instruments we used the FoV sizes and effective areas $A_{\rm eff}$  from the instruments' manuals and websites.

The results of this calculation are shown in Fig.~\ref{fig8} by straight lines. Each line shows the locus of the points on the $f_{\rm sky}$--$A_{\rm eff}\times t$ plane, which can be achieved in the course of a one-year survey. For eROSITA we also did a calculation for 4 years. Different locations along these lines correspond to different fractions of the sky covered in the survey (and to different survey depth, as the total duration of the survey was fixed). Obviously,  the optimal survey parameters are defined by the point where the signal-to-noise ratio is maximal. Owing to the shape of the iso-contours on the signal-to-noise map, the optimum is not achieved by covering the whole sky. On the contrary, for the low-grasp missions, it is reached by surveying rather small sky areas, $\sim 40-400$ deg$^2$ to large depths in order to collect a large number of counts. In this context it is worth mentioning that, as we study CXB surface brightness fluctuations, considerations  of the confusion limit are irrelevant.

From Fig.\ref{fig8} one can see that to detect the tomographic signal at the confidence level of $\ga 100 \sigma$ a survey with a $\sim 10$ m$^2$ class instrument is required. Such a detection would permit detailed redshift-resolved studies of the correlation properties of AGN in the $z\sim 0-2$ redshift range. 

Among planned and proposed missions carrying X-ray optics, the highest signal-to-noise ratio of $\sim 25-40\sigma$ can be achieved by WFXT in a survey covering $\sim 3\cdot 10^3$ deg$^2$ ($\sim 7\cdot 10^3$ deg$^2$ for a 500 eV EW line). In the all-sky survey of the same duration, WFXT will achieve $\sim 15-30\sigma$ detection. With somewhat lower confidence the signal will be measured by an ATHENA class mission. 

eROSITA could in principle detect the signal at the $\sim 12\sigma$ confidence level ($\sim 20\sigma$ for a 500 eV EW line), if the four-year survey concentrated on the  $\sim 700$ deg$^2$ ($\sim 2000$ deg$^2$) region of the sky, which is unrealistic to expect, as such a ``pencil beam'' survey would undermine the main scientific objectives of the mission. In the all-sky survey eROSITA will detect the tomographic signal only marginally, if at all.

It is worth noting that the measurement of the tomographic  signal describing intensity fluctuations in the line emission ($C_{\ell}$ in the terminology of Eq.~(\ref{eq13})) should not be confused with the detection of CXB intensity fluctuations due to continuum emission of objects located at all redshifts ($C_{\ell}^{(11,22)}$ in Eq.~(\ref{eq13})). The power spectrum of the latter can be detected by many missions, including eROSITA \citep{kolodzig_II}. 

Among currently operating missions, only XMM-Newton has a chance of a $\sim 8\sigma$ ($\sim 13\sigma$ for a 500 eV EW line) detection of the tomographic signal. This would require a one-year long survey covering $\sim 150$ deg$^2$ ($\sim 400$ deg$^2$ for a 500 eV EW line) of the sky, with the exposure time of $\sim 40$ ksec ($\sim 15$ ksec) per pointing. Although this may sound like an enormous investment of observing time, it is not entirely unfeasible, given the long lifetime of the mission and the breadth of the science topics which may be addressed by such a survey.

\begin{figure*}
\centering
\includegraphics[width=\plotwdtwo]{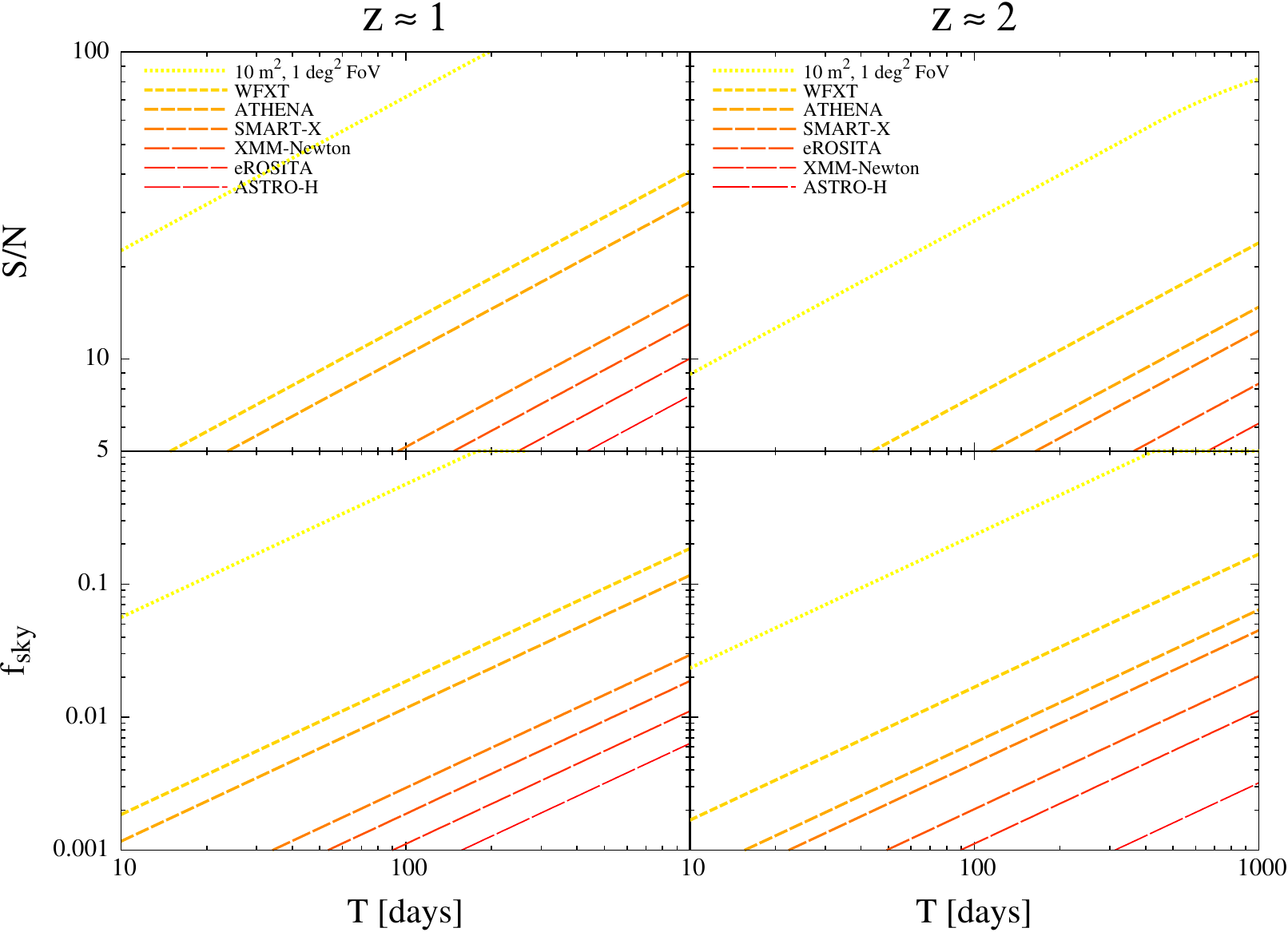}
\caption{Achievable $S/N$ and optimal  sky fraction covered by the survey  as a function of total survey time for various existing and future X-ray instruments. Calculations were done for Model I. Left- and right-hand panels correspond to $3.0-3.5$ keV and $1.6-2.1$ keV observational energy ranges. These ranges are tuned to detect the tomographic signal from the redshift $z\simeq 1$ and $z\simeq 2$, respectively.}
\label{fig9}
\end{figure*}

In Fig.~\ref{fig9} we plot achievable signal-to-noise ratio and the optimal survey sky fraction (assuming Model I) as a function of total survey time for various instruments. Because grazing incidence telescopes often have a significant jump in the effective area below $E \sim 2.1-2.2$ keV, we also did these calculations for the $E^{(1)}=1.6-1.8$ keV and $E^{(2)}=1.9-2.1$ keV energy bins, corresponding to the redshift $z\simeq 2$, the result plotted  in the right-hand panels of Fig.~\ref{fig9}. 
However,  no significant gain  is achieved at lower energies, with  the signal-to-noise ratio even decreasing somewhat. The main reason is the increased cosmological dimming along with reduced volume factor at higher redshifts, which cannot be compensated for by about a four- to five-fold increase in the effective area.

From Fig.~\ref{fig9} we see that the optimal survey strategy requires that $f_{\rm sky}\propto T$ with the proportionality coefficient determined by the grasp of the instrument. This can be understood in terms of the competition between the photon-counting noise and the AGN discreteness noise. Indeed, the time $t$ spent on an individual pointing should be large enough to reduce the photon-counting noise to the level comparable to the (irreducible) AGN discreteness noise. Any further increase in the exposure time $t$ does not result in any significant increase in the signal-to-noise ratio. Therefore, irrespective of the total survey time $T$, the optimal time $t$ spent per pointing is fixed for a given instrument. A consequence of this is that the signal-to-noise ratio scales with the survey time as $S/N\propto\sqrt{f_{\rm sky}}\propto\sqrt{T}$.

For X-ray instruments there is usually a tradeoff between angular resolution and effective area. In this context it is important to realize that the peak of the signal-to-noise ratio is achieved at the angular scales corresponding to $\ell\sim 100-300$ (Fig.~\ref{fig3},\ref{fig7}), therefore  moderate angular resolution of $\sim 0.1-0.3$ degrees or even coarser is sufficient. Therefore the main limiting factor for the present and near-term missions is the grasp, $A_{\rm eff}\times{\rm FoV}$, but not the angular resolution. As it turns out, among currently discussed missions, the one with the highest grasp is LOFT (Large Observatory For X-ray Timing)~\footnote{\url{http://gri.rm.iasf.cnr.it/}}, even though it was proposed for an entirely different purpose and does not carry X-ray optics. With the proposed effective area of $\sim 10$ m$^2$, it is potentially capable of detecting the tomographic signal with the signal-to-noise ratio of $\sim 100\sigma$. For LOFT detector, the angular resolution of the sky intensity map will be determined by its  field of view, which  is expected to be in the $\sim 0.5-1$ deg range. This corresponds to $\ell \sim 200-400$, and this level of angular smearing should not lead to a significant deterioration in the signal-to-noise ratio (Fig.~\ref{fig3}).  The main potential obstacle in using LOFT for the $6.4$ keV line tomography (apart from the observation planning considerations) is the amplitude and stability of the instrumental background.

\section{Discussion and conclusions}\label{sec4}

Our results show that the iron $6.4$ keV line tomography of LSS should  be possible with a survey covering a considerable portion of the sky with the sensitivity that allows detection of  $\sim 500-1000$ counts from a  $10^{-13}$ erg/s/cm$^2$ source ($2-10$ keV band). Such sensitivity corresponds to a $\sim 25 $ ksec XMM-Newton observation (PN+2MOS detectors).

This 6.4  keV line tomography can complement and compete with the more traditional methods of studying LSS via  building source catalogs and analyzing the 3D distributions of sources.
Its main advantage over the traditional methods is that no redshift information is required, thus alleviating the need for optical follow-up observations. Such observations may be especially time consuming for faint sources. Moreover, they may become prohibitively long when large sky areas are considered. 

Since the goal of the intensity mapping is not to resolve the fluctuation field down to all the discrete components, but rather to investigate the statistics of fluctuations on somewhat larger scales, the issue of source confusion is  irrelevant. Therefore  no demanding requirements are imposed  on the angular resolution of the instrument. Indeed, the major contribution to the signal-to-noise ratio is made by angular scales corresponding to $\ell \sim 100-300$ (Fig.~\ref{fig3},~\ref{fig7}), i.e. $\sim 0.5-2\deg$. This is also comfortably consistent with our calculations using only information from the scales, where the assumptions of linear evolution and Gaussianity are very justified. The corresponding maximal multipole number is $\ell_{\max}\approx 500$. Typical angular resolutions of modern X-ray telescopes are much better than these angular scales, therefore there is no need to include the effect of the instrumental point spread function in our calculations.

The iron K$\alpha$ line tomographic signal is sensitive to the effective AGN clustering bias, AGN LF, and $6.4$ keV line parameters. From the very detailed, deep but narrow-field, X-ray studies one could get a good handle on AGN LF along with estimates for the population-averaged $6.4$ keV line strength, and apply this knowledge in tomographic measurements to determine AGN clustering bias as a function of redshift. On the other hand, even the best currently available X-ray AGN LFs are based on a rather limited number of objects, $\sim 10^3$, primarily  detected  in a few narrow, pencil-beam surveys, therefore become progressively less accurate with increasing redshift and luminosity. Also, they may be subject to the cosmic variance. This may affect our predictions for the strength of the tomographic signal to be measured  in large area  surveys and at lower energies, corresponding to higher redshifts.  Therefore actual detection and measurement of the tomographic signal can help constrain the evolution of the AGN volume density, up to  $z\sim 2$, and maybe slightly beyond, provided that reasonable assumptions about the redshift behavior of the AGN bias are made.

 On the other hand, if one has a good empirical model for the AGN clustering and LF available, one could turn the above argument around to learn something about the population-averaged $6.4$ keV line strength and its possible evolution with redshift. This gives us a probe of Compton thick fraction of AGN and its evolution over cosmic time.

In our calculations, we used a simple spectral model, consisting of a power-law continuum and a narrow line.   Real AGN spectra are more complex. First, the  iron $6.4$ keV line has also a broad component, which has an intermediate behavior in terms of variation as a function of energy.  Second, in the energy range of interest there are weaker lines, including the $6.7$ and $6.9$ keV lines of He,H-like iron, present in some of the AGN spectra and in the spectra of clusters of galaxies. Among other lines are the iron $K_\beta$ fluorescent  line at $7.06$ keV and the Ni $K_\alpha$ line at  $7.5$ keV. These lines are $\ga 5-10$ times weaker than the $6.4$ keV line, and although they should be taken into account in  more precise calculations, their contribution was ignored in our study.

Third, the shape of the continuum spectrum is more complex than a power law. The most important feature is the iron K-edge at $7.1$ keV. As the fluorescent yield of iron $K_\alpha$ line is $\approx 0.3$ ($\approx 0.038$ for the $K_\beta$ line) \citep{1972RvMP...44..716B,1974A&A....31..249B}, the  K-edge is produced by removing about three times more photons from the spectrum than contained  in the $6.4$ keV line. However, in the case of the reflected spectrum, these photons are distributed over a significantly  broader energy interval than the narrow $6.4$ keV line, and the amplitude of the resulting feature at $7.1$ keV is correspondingly smaller.  
Similar to the line, the K-edge has two components, a narrow component with a sharp step-like feature at $7.1$ keV and the relativistically broadened component, also known as the ``smeared edge''. The relative strength of the two components will vary according to the relative strengths of broad and narrow line components.  Since the location and depth of the sharp step-like feature at $7.1$ keV are defined by the laws of atomic physics and are precisely known, one can use a spectral template that includes both the line and the edge,  in measuring the tomographic signal.  This  can potentially increase the signal-to-noise ratio of the tomographic signal. Similarly, other lines like the iron $K_\beta$ and nickel $K_\alpha$ lines, can be included in such a template, resulting in a further increase in the signal-to-noise ratio.

The complexities described above were ignored in our study, because its main goal  was  to obtain a simple, hence inevitably somewhat rough, estimate of the observability of the tomographic signal.  We thus have intentionally kept our model as simple as possible and considered a spectrum consisting of two components with clearly distinct behavior as a function of energy -- slowly varying continuum and rapidly changing narrow line.
Obviously, this analysis  can be easily extended to incorporate more complex AGN spectra. 
Below, we very briefly sketch the way one might proceed by using the Fisher matrix approach \citetext{see, e.g., \citealp{1997ApJ...480...22T}}.

The Fisher information matrix (i.e., the ensemble average of the Hessian matrix of the minus log-likelihood) for the CXB fluctuation fields measured in energy bins $i$ and $j$ (and assuming Gaussianity) can be given as
\begin{equation}
F^{(ij)}_{mn}=\frac{1}{\left(\delta C_{\ell_m}^{(ij)}\right)^2}\delta_{\ell_m\ell_n}\,,
\end{equation}
where $\delta C_{\ell_m}^{(ij)}$ is given by Eq.~(\ref{eq17}), and $\delta_{mn}$ denotes Kronecker delta. Here the parameterization is in the form of discrete bandpowers $C_{\ell_m}^{(ij)}$. If $C_{\ell}^{(ij)}$ is fully determined by parameter vector ${\bf \Theta}=\{\theta_{\alpha}\}$, $\alpha=1\ldots N$, $F^{(ij)}_{mn}$ can be rotated into that basis, giving
\begin{eqnarray}
F^{(ij)}_{\alpha\beta}&=&\sum\limits_m\sum\limits_n\frac{\partial C_{\ell_m}^{(ij)}}{\partial \theta_{\alpha}}F^{(ij)}_{mn}\frac{\partial C_{\ell_n}^{(ij)}}{\partial \theta_{\beta}}=\nonumber\\
&=&\sum\limits_{\ell}\frac{\partial C_{\ell}^{(ij)}}{\partial \theta_{\alpha}}\frac{1}{\left(\delta C_{\ell}^{(ij)}\right)^2}\frac{\partial C_{\ell}^{(ij)}}{\partial \theta_{\beta}}\,.
\end{eqnarray}
Here the parameter vector could contain, e.g., parameters of the AGN spectral template, clustering bias parameters, and parameters describing AGN LF. Since the observations can be done in several frequency bins, the total Fisher information matrix can be written as
\begin{equation}
F_{\alpha\beta}=\sum\limits_i\sum\limits_{j\le i}F^{(ij)}_{\alpha\beta}\,.
\end{equation}
Here $j\le i$ is to ensure that cross-bin contributions are included only once. Once $F_{\alpha\beta}$ is calculated, the obtainable parameter constraints follow immediately \citetext{see, e.g., \citealp{1997ApJ...480...22T}}. We leave the detailed implementation of the above scheme to a possible, future paper.

It is also important to point out that, even though the analysis in this paper focused on large-scale clustering signal, and thus assumed significant sky coverage, the method is also applicable to smaller survey fields, where one can only effectively probe the one-halo term. In this case the variability of the amplitude of the one-halo term as a function of energy should provide one with means to probe the flux-weighted number density of AGN as a function of redshift.

Finally, although in this paper we did not discuss the possibility of cross-correlating CXB maps at different energy ranges with the (photometric or spectroscopic) galaxy catalogs, it is certainly one of the ways to enhance the fidelity of the tomographic signal.

We conclude that the $6.4$ keV line tomography of the LSS is indeed feasible with the future X-ray instruments.  In particular, WFXT/ATHENA type missions should be able to detect the tomographic signal with a moderate significance, whereas a 10 m$^2$ class mission will perform detailed tomography of the LSS. The LOFT detectors, although designed for entirely different  science goals and not equipped with X-ray optics, have the largest grasp among currently operating, planned or proposed missions, in the same range of values  as  our hypothetical 10 m$^2$ case. Therefore LOFT has the potential to detect the tomographic signal with a high signal-to-noise ratio. A more detailed feasibility study should take the amplitude and stability of the instrumental background of LOFT detectors into account. Furthermore, for this potential to be realized, a dedicated effort should be made to accommodate at least a few month-long sky survey in the LOFT observing program.

\acknowledgements{We thank our referee for comments and suggestions. GH thanks Rishi Khatri for useful discussions.}
\bibliographystyle{aa}
\bibliography{references}

\end{document}